\documentclass[a4paper, conference]{IEEEtran}
\IEEEoverridecommandlockouts
\usepackage{cite}
\usepackage{amsmath,amssymb,amsfonts}
\usepackage{algorithmic}
\usepackage{graphicx}
\usepackage{textcomp}
\usepackage{xcolor}
\usepackage{url}
\usepackage[absolute,overlay]{textpos}

\def\BibTeX{{\rm B\kern-.05em{\sc i\kern-.025em b}\kern-.08em
    T\kern-.1667em\lower.7ex\hbox{E}\kern-.125emX}}
\begin{document}

\title{Time-domain wideband image source method for spherical microphone arrays\\
\thanks{The authors would like to thank Prof. Dr.-Ing. Sascha Spors, Dr.-Ing. Frank Schultz and Nara Hahn for discussing relevant concepts.}
}

\author{\IEEEauthorblockN{Jiarui Wang}
\IEEEauthorblockA{\textit{The Australian National University} \\
Canberra, Australia\\
u5879960@anu.edu.au}
\and
\IEEEauthorblockN{Jihui Aimee Zhang}
\IEEEauthorblockA{\textit{University of Southampton} \\
Southampton, UK\\
Aimee\_Jihui.Zhang@soton.ac.uk}
\and\IEEEauthorblockN{Prasanga Samarasinghe, Thushara Abhayapala}
\IEEEauthorblockA{\textit{The Australian National University} \\
Canberra, Australia\\
prasanga.samarasinghe@anu.edu.au}
}

\maketitle

\begin{textblock*}{18cm}(1cm,26cm) % {block width} (coords) 
  \fbox{\begin{minipage}[c]{\textwidth}
        \centering\small{ \copyright 2023 IEEE. Personal use of this material is permitted. Permission 
from IEEE must be obtained for all other uses, in any current or future 
media, including reprinting/republishing this material for advertising or 
promotional purposes, creating new collective works, for resale or 
redistribution to servers or lists, or reuse of any copyrighted 
component of this work in other works.}\end{minipage}}
\end{textblock*}

\begin{abstract}
This paper presents the time-domain wideband spherical microphone array impulse response generator (TDW-SMIR generator), which is a time-domain wideband image source method (ISM) for generating the room impulse responses captured by an open spherical microphone array. To incorporate loudspeaker directivity, the TDW-SMIR generator considers a source that emits a sequence of spherical wave fronts whose amplitudes are related to the loudspeaker directional impulse responses measured in the far-field. The TDW-SMIR generator uses geometric models to derive the time-domain signals recorded by the spherical microphone array. Comparisons are made with frequency-domain single band ISMs. Simulation results prove the results of the TDW-SMIR generator are similar to those of frequency-domain single band ISMs. 
\end{abstract}


%%%%%%%%%%%%%%%%%%%%%%%%% Introduction %%%%%%%%%%%%%%%%%%%
\section{Introduction}
Image source method (ISM) is widely used in the simulation of room acoustics. Room impulse responses generated by ISMs can be used in perceptual tests \cite{SporsWFSEval, Lee2016} and generating training datasets \cite{ADIWAENC, ADIS}. Traditionally, ISM concerns the point-to-point room impulse response \cite{BerkleyISM}, which can be transformed into the frequency-domain to form the point-to-point room transfer function. Variants of the traditional ISM exist \cite{DuraiswamiISM, McGovern2009, LehmannISM}, which mainly focus on further reducing computational complexity. 

Spherical microphone array captures the sound field in the three dimensional space \cite{Meyer2002, Thushara2002, Rafaely2005}. Due to the recent interest in spatial sound field reproduction and virtual reality, simulating the room impulse responses captured by a spherical microphone array becomes increasingly important.  Four examples of spherical microphone array room impulse response generator are \cite{JarrettSMIR},  \cite{Luo2021}, \cite{Naylor2015} and \cite{PrasangaISM}. In this paper, all four examples are called single band ISM. This is because in their procedures, the room transfer functions are first calculated at individual frequency bins, followed by the inverse Fourier transform to obtain the time-domain room impulse responses. The procedures are different from wideband or time-domain approaches where a broad frequency range is considered together. Room impulse response simulation should also consider source directivity. Numerous source directivity datasets exist \cite{Brandner2018, Tylka2015, Shabtai2017}. So far, most ISMs incorporated source directivity in the frequency-domain \cite{JarrettSMIR, Luo2021, Naylor2015, PrasangaISM, Bu2017}. Single band ISMs for spherical microphone arrays such as \cite{JarrettSMIR} and \cite{PrasangaISM} involve the calculation of the spherical Bessel function and the spherical Hankel function, which could be computationally intensive. 

In this paper, ISMs that directly model the room impulse responses in the time-domain without first calculating the frequency-domain room transfer functions are called time-domain wideband ISMs. In time-domain wideband ISMs, the delays of the delta functions are directly computed, which allows the direct control of causality. The original ISM in \cite{BerkleyISM} is a time-domain wideband ISM. The delays of the delta functions can be modelled by low-pass impulse method \cite{Peterson1986}. In the time-domain, the source directivity is characterized by the directional impulse responses, which is the counterpart of the commonly used frequency-domain directivity pattern. The direct path and the reflections can be filtered by the directional impulse responses in the time-domain to include the source directivity, which is most suitable for a single receiver. When the receiver is a spherical microphone array, repetitive calculations are performed for each microphone. Spherical harmonic (SH) functions are widely used to model signals on a spherical surface \cite{Kennedy2013}, which enables the simultaneous calculation of the impulse responses  at all microphone locations. Moreover, because source directivity is usually measured on a spherical measurement surface with fixed radius, it can also be expressed using SH functions. Finite-difference time-domain (FDTD) methods can also incorporate source directivity and directly provide the time-domain solutions of the room impulse responses\cite{Murphy2014, Takeuchi2019}. However, due to the requirement of specialized software and large computational resources often associated with FDTD methods, this paper restricts the scope to ISMs.

This paper presents the time-domain wideband spherical microphone array impulse response generator (TDW-SMIR generator). The TDW-SMIR generator is based on the time-domain wideband ISM in \cite{Wang2023} and simulates the room impulse responses captured by an open spherical microphone array directly in the time-domain using SH functions. The room is restricted to cuboid shape. The TDW-SMIR generator allows loudspeaker directivities that are of arbitrary shape and frequency dependent. To incorporate loudspeaker directivity, the source in the TDW-SMIR generator emits a sequence of spherical wave fronts whose amplitudes are related to the measured far-field loudspeaker directional impulse responses. The amplitudes of the spherical wave fronts are expressed using SH functions. The method presented in this paper can also be used to simulate the time-domain signals observed on the boundary of a spherical listening region as in \cite{Wang2023}. Due to the space limit, detailed evaluations were not presented in \cite{Wang2023}. In this paper, comparisons of the TDW-SMIR generator and two single band ISMs for spherical microphone arrays are presented. In Section \ref{Sec:SMIR_compare}, the TDW-SMIR generator is compared with the SMIR generator in \cite{JarrettSMIR}, which is available open source in \cite{SMIRWeb}. The comparison focuses on sources with frequency invariant cardioid-type directivity. In Section \ref{Sec:R2R_RTF_compare}, the TDW-SMIR generator is compared with the region-to-region ISM in \cite{PrasangaISM}, which can be adapted to simulate the signals captured by a spherical microphone array in the frequency-domain. The comparison focuses on sources with frequency dependent and arbitrarily shaped directivity. The method in \cite{PrasangaISM} relies on the addition theorem for the spherical Hankel function, which is not utilized in the TDW-SMIR generator.

The TDW-SMIR generator in this paper only considers open spherical microphone array. Time-domain wideband simulation of signals captured by a rigid spherical microphone array in the free-field is covered in \cite{HahnRigid21} and \cite{HahnRigid23}. Both \cite{HahnRigid21} and \cite{HahnRigid23} focused on pole-zero analysis of the transfer function. The application of geometric models to ISM for rigid spherical microphone arrays can be considered in future work. 

Future work should also investigate the perceptual effects of the TDW-SMIR generator. Because the TDW-SMIR generator is for rooms with perfect cuboid geometry and relies on reflecting sources across rigid walls, the resulting room impulse responses is likely to suffer from sweeping echo problem \cite{KiyoharaSE}. Methods to reduce sweeping echos are covered in \cite{DSSE} and \cite{MGSE}. Moreover, the order (number) of the image sources could be reduced and a perceptually inspired way should be used to generate the late reverberation. 

%%%%%%%%%%%%%%%%%%%%%%%% Summary of ISM %%%%%%%%%%%%%%%%%
\section{Summary of the time-domain wideband ISM}
%%%%%%%%%%%%%%%%%%%%%%%%% Anechoic %%%%%%%%%%%%%%%%%%%%%%
\subsection{Anechoic case}
\label{subsec:anechoic}
\begin{figure}[t]
  \centering
  \includegraphics[trim = 4mm 20mm 2mm 18mm, clip, width = 0.9\columnwidth]{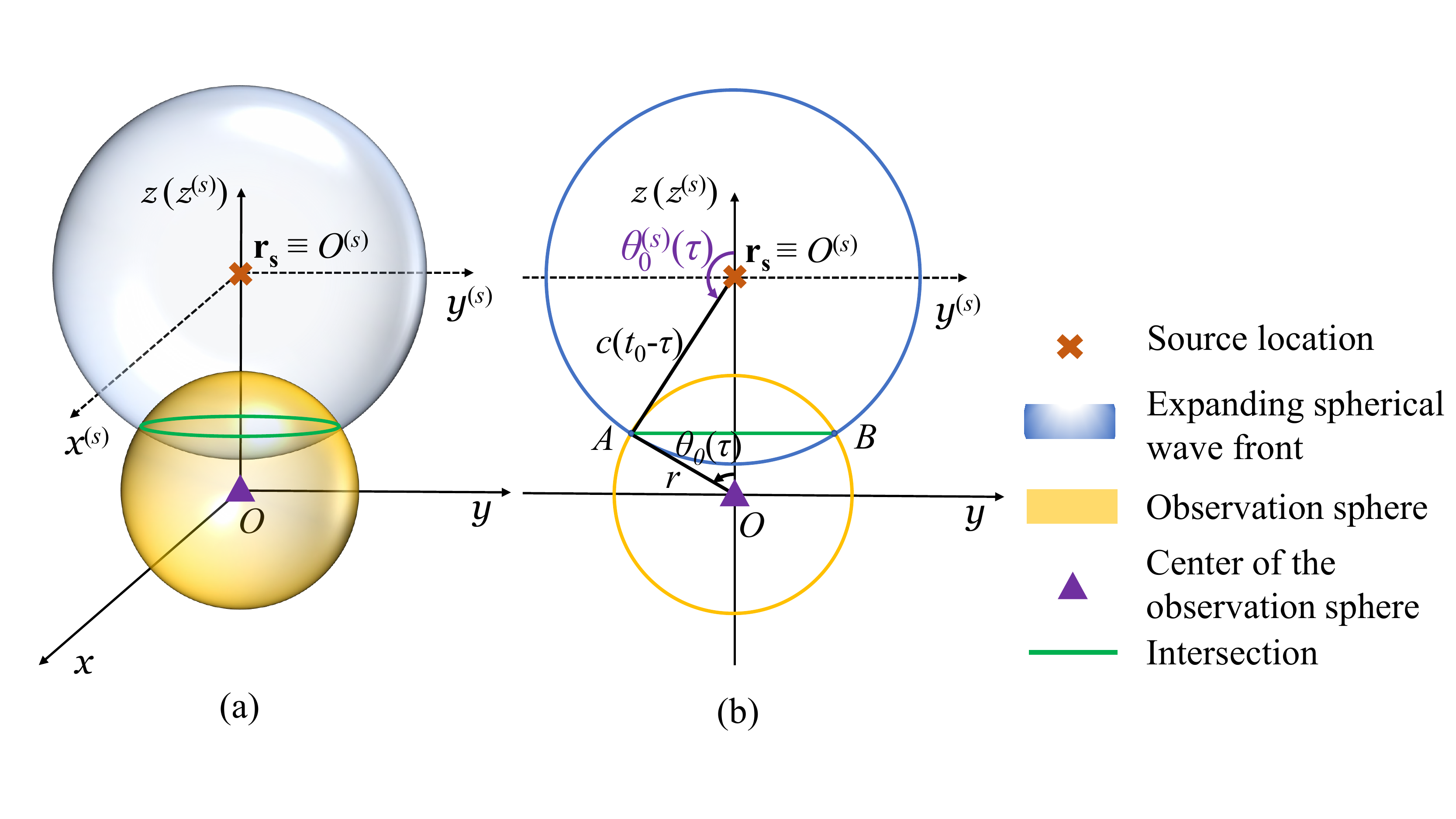}
  \vskip -3mm
  \caption{(a) 3D view and (b) cross section view of the observation sphere in yellow and the expanding spherical wave front in light blue. The source is at $\mathbf{r_{s}}$. Line $AB$ is the cross section of the circle, which is the intersection of the observation sphere and the expanding spherical wave front. }
  \vskip -3mm
  \label{Fig:cross_section}
\end{figure}
Figure \ref{Fig:cross_section} shows the setup, which was also used in \cite{Wang2023} and \cite{Wang2022} to derive the observed time-domain signal on the observation sphere due to a source that emits a sequence of spherical wave fronts in the free-field. The observation sphere could be the boundary of the spherical listening region in the sound field reproduction system, or the boundary of the open spherical microphone array as in this paper. Similar geometric models appeared in \cite{Jensen1999} and \cite{Poletti2016}. The observation sphere is of radius $r$. The source $\mathbf{r_{s}}$ is on the positive $z$-axis. In this paper, the superscript shows the coordinate system used to express the location. If there are no superscripts, then the location is expressed with respect to the $xyz$ coordinate system.

The source emits a sequence of spherical wave fronts 
\begin{align}
\label{Eq:source_seq}
    d(t, \theta^{(s)}, \phi^{(s)}) &= \int_{\tau} \widehat{d}(\tau, \theta^{(s)}, \phi^{(s)}) \delta(t-\tau) d\tau \nonumber\\ &= \int_{\tau} \sum_{v = 0}^{V} \sum_{u = -v}^{v} \gamma_{v}^{u}(\tau) Y_{v}^{u}(\theta^{(s)}, \phi^{(s)})\delta(t - \tau)d\tau 
\end{align}
in which $V$ is the SH truncation order. In the free-field, each spherical wave front expands at $c$ meters per second. For the spherical wave front emitted at $t =\tau$, when $t \in [(r_{s}-r)/c+\tau, (r_{s}+r)/c+\tau]$, it intersects the observation sphere. The intersection is a circle parallel to the $xy$ plane. Suppose at $t = t_{0}$, the circle is at elevation angle $\theta = \theta_{0}(\tau)$ (equivalent to $\theta^{(s)} = \theta_{0}^{(s)}(\tau)$). From \cite{Wang2023}, considering the sequence of spherical wave fronts in \eqref{Eq:source_seq}, the observed signal at $t=t_{0}$ is 
\begin{align}
\label{Eq:obv_WF_sym}
    g(t_{0}, r, \theta, \phi) = &\int_{\tau}\frac{c}{4\pi r r_{s}} \delta \big(\cos \theta - \cos \theta_{0}(\tau)\big) \nonumber\\ &\widehat{d}\big(\tau, \theta_{0}^{(s)}(\tau), \phi\big) d\tau
\end{align}
in which $\phi \equiv \phi^{(s)}$. Equation \eqref{Eq:obv_WF_sym} is the superposition of multiple circles, where each circle has direction dependent amplitude determined by the term $\widehat{d}\big(\tau, \theta_{0}^{(s)}(\tau), \phi)$.
Let $\zeta_{n}^{m}(t_{0}, r)$ denote the SH coefficients of \eqref{Eq:obv_WF_sym}. According to \cite{Wang2023},
\begin{align}
\label{Eq:SH_coeff_obv_seq_gen}
    \zeta_{n}^{m}(t_{0}, r) &= \frac{c}{2 r r_{s}} \int_{\tau} \sum_{v=0}^{V} \sum_{u = -v}^{v} \gamma_{v}^{u}(\tau)\; \mathcal{P}_{v}^{u}[\cos\theta_{0}^{(s)}(\tau)] \nonumber\\  &\mathcal{P}_{n}^{m}[\cos\theta_{0}(\tau)] \delta_{m, u} \;\Xi(t_{0}, \tau) \;d\tau
\end{align}
in which $\delta_{m, u}$ is the Kronecker delta function, and $\mathcal{P}_{n}^{m}(\cdot) = \sqrt{(2n+1)/{4\pi}}\sqrt{(n-m)!/(n+m)!} P_{n}^{m}(\cdot)$ with $P_{n}^{m}(\cdot)$ denoting the associated Legendre polynomial of degree $n$ and order $m$. In practice, the SH coefficients $\zeta_{n}^{m}(t_{0}, r)$ are truncated to degree $n=N$. From Figure \ref{Fig:cross_section}, the arguments
\begin{equation}
\label{Eq:cos_theta_seq}
    \cos \theta_{0}(\tau) = \frac{r^{2} + r_{s}^{2} - c^{2}(t_{0}-\tau)^{2}}{2rr_{s}} ,
\end{equation}
\begin{equation}
\label{Eq:cos_vartheta_seq}
    \cos\theta_{0}^{(s)}(\tau) = -\frac{c^{2}(t_{0}-\tau)^{2} + r_{s}^{2} - r^{2}}{2c(t_{0}-\tau) \,r_{s}} .
\end{equation}
The rectangular window $\Xi(t_{0}, \tau) = 1$ when $r_{s}-r \leq c(t_{0}-\tau) \leq r_{s}+r$; otherwise, $\Xi(t_{0}, \tau) = 0$. For each $v, u$ pair, equation \eqref{Eq:SH_coeff_obv_seq_gen} can be treated as a convolution integral where one function is $\gamma_{v}^{u}(t_{0})$ and the other function is $\mathcal{P}_{v}^{u}[\cos\theta_{0}^{(s)}(0)]\mathcal{P}_{n}^{m}[\cos\theta_{0}(0)] \delta_{m, u} \;\Xi(t_{0}, 0) $.

%%%%%%%%%%%%%%%%%%%%%%% The proposed ISM %%%%%%%%%%%%%%%%%%%
\subsection{The time-domain wideband ISM}
\label{subsec:ISM_steps}
\begin{figure}[t]
  \centering
  \includegraphics[trim = 70mm 15mm 70mm 5mm, clip, width = 0.9\columnwidth]{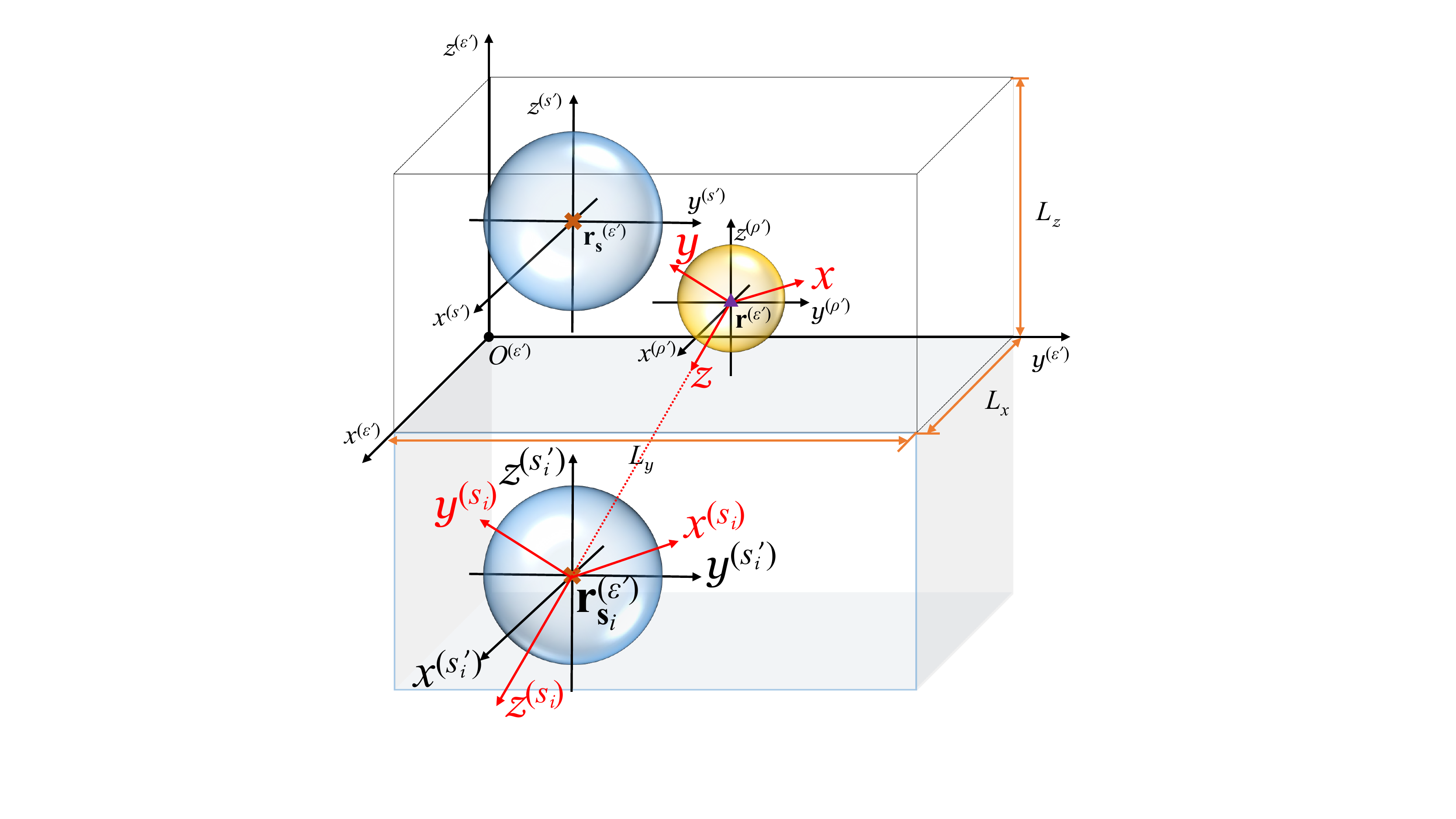}
  \vskip -3mm
  \caption{Setup of the image source method. The original source is at  $\mathbf{r_{s}^{(\varepsilon')}}$. The image source at $\mathbf{r_{s_{\mathit{i}}}^{(\varepsilon')}}$ is the mirror reflection of the original source through the $x^{(\varepsilon')}y^{(\varepsilon')}$ plane. The observation sphere in yellow is centered at $\mathbf{r^{(\varepsilon')}}$. The mirrored room is shaded. }
  \vskip -3mm
  \label{Fig:ISM_room_setup}
\end{figure}
Like traditional ISMs \cite{BerkleyISM}, the time-domain wideband ISM in \cite{Wang2023} replaces wall reflections by image sources located in a grid of mirrored rooms. Figure \ref{Fig:ISM_room_setup} shows the setup. The original source is located at $\mathbf{r_{s}^{(\varepsilon')}}$ and the observation sphere is centered at $\mathbf{r^{(\varepsilon')}}$. The image source $\mathbf{r_{s_{\mathit{i}}}^{(\varepsilon')}}$ formed by reflecting the original source through the $x^{(\varepsilon')}y^{(\varepsilon')}$ plane is also shown. For each image source, the amplitudes of the emitted spherical wave fronts are the mirror reflections of those emitted by the original source. To use the results in Section \ref{subsec:anechoic}, the red coordinate systems in Figure \ref{Fig:ISM_room_setup} are introduced so that the image source is on the positive $z$-axis. Note that the orientations of the $xyz$ coordinate system and the $x^{(s_{i})}y^{(s_{i})}z^{(s_{i})}$ coordinate system are unique to each image source. The black coordinate systems $x^{(s')}y^{(s')}z^{(s')}$, $x^{(s'_{i})}y^{(s'_{i})}z^{(s'_{i})}$ and $x^{(\rho')}y^{(\rho')}z^{(\rho')}$ have the same orientation as the room coordinate system $x^{(\varepsilon')}y^{(\varepsilon')}z^{(\varepsilon')}$. They are convenient to use when expressing the amplitudes of the emitted spherical wave fronts and the observed signal. 

Suppose the source emits a sequence of spherical wave fronts $d(t, \theta^{(s')}, \phi^{(s')})$. \textbf{First}, the SH coefficients of the reflected spherical wave fronts $d(t, \theta^{(s_{i}')}, \phi^{(s_{i}')})$ emitted by the image sources can be calculated by using the parity properties of SH functions. Examples of the parity properties related to the ISM can be found in \cite{JarrettSMIR, PrasangaISM, Bu2017} and \cite{Wang2023}. In the \textbf{second} step, a rotation of the coordinate system is performed so that $d(t, \theta^{(s_{i}')}, \phi^{(s_{i}')})$ is converted to $d(t, \theta^{(s_{i})}, \phi^{(s_{i})})$, which represents the amplitudes of the spherical wave fronts emitted by the image source expressed with respect to the $x^{(s_{i})}y^{(s_{i})}z^{(s_{i})}$ coordinate system. The rotation can be achieved by using the Wigner $D$-matrix \cite{Kennedy2013} in the SH domain. In the \textbf{third} step, the SH coefficients of the observed signal with respect to the $xyz$ coordinate system are calculated by following Section \ref{subsec:anechoic}. The attenuation due to wall reflections is also considered in this step. In the \textbf{fourth} step, another rotation of the coordinate system is performed so that the SH coefficients of the observed signal are expressed with respect to the $x^{(\rho')}y^{(\rho')}z^{(\rho')}$ coordinate system. The Wigner $D$-matrix can be used to perform rotation in the SH domain. This step is necessary because a common coordinate system must be chosen before summing the contributions of different image sources. \textbf{Finally}, the contributions of all image sources are added together. Mathematical details are covered in \cite{Wang2023}.

%%%%%%%%%%%%%%%%%%% Comparison with SMIR %%%%%%%%%%%%%%%%%%%%%
\section{The TDW-SMIR generator with frequency invariant directional sources}
\label{Sec:SMIR_compare}
%%%%%%%%%%%%%%%%%%%%%%%%%%%%%%%%%%%%%%%%%%%%%%%%%%%%%%%%%%%%%
\begin{figure}[t]
  \centering
  \includegraphics[trim = 40mm 25mm 40mm 30mm, clip, width = 0.85\columnwidth]{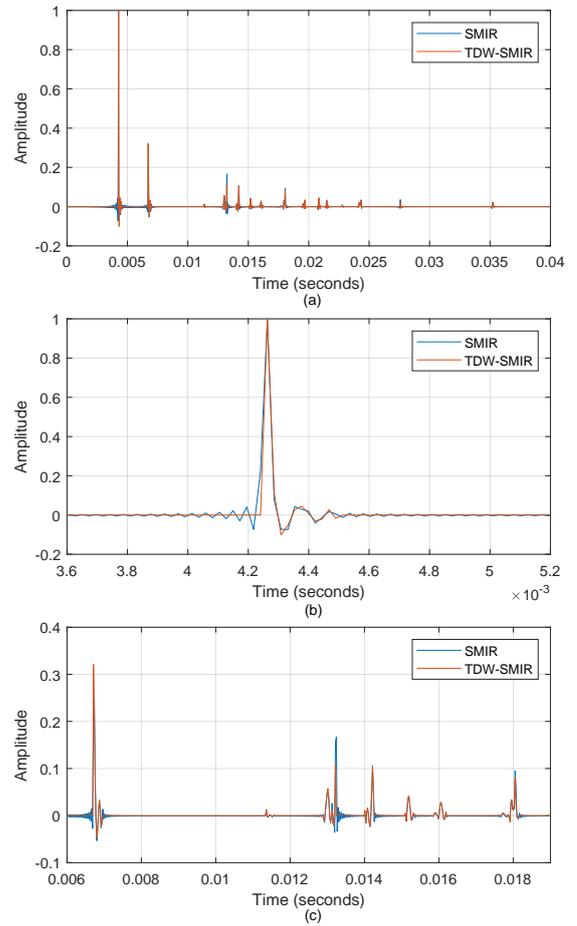}
  \vskip -3mm
  \caption{Comparison of the impulse responses at microphone No.17 obtained using the SMIR generator and the TDW-SMIR generator. (a) The whole impulse response. (b) The direct path. (c) A few early reflections. }
  \vskip -3mm
  \label{Fig:ISM_cardioid_results}
\end{figure}
The TDW-SMIR generator can be applied to sources whose directivity is frequency invariant, i.e., $d(\theta^{(s')}, \phi^{(s')}, f) = \widehat{d}(\theta^{(s')}, \phi^{(s')}) \;\forall f$. Sources with frequency invariant directivity are featured in the SMIR generator in \cite{JarrettSMIR}. The SMIR generator is a frequency-domain single band method that calculates the room transfer functions captured by a spherical microphone array. Inverse Fourier transform is then applied to obtain the room impulse responses. Simple source directivities are considered including omnidirectional, cardioid, hypercardioid, subcardioid and bidirectional patterns. The room transfer functions are formulated in the SH domain. 

Here we propose the TDW-SMIR generator for sources with frequency invariant directivity. Taking the inverse Fourier transform of the source directivity, $d(t, \theta^{(s')}, \phi^{(s')}) = \widehat{d}(\theta^{(s')}, \phi^{(s')}) \delta(t)$ (ignoring the $2\pi$ scaling factor). Compared with \eqref{Eq:source_seq}, this is equivalent to a source that emits a spherical wave front whose amplitude is $\widehat{d}(\theta^{(s')}, \phi^{(s')})$ at  $t=0$ seconds. The TDW-SMIR generator then follows the ISM in Section \ref{subsec:ISM_steps}. Compared with the SMIR generator in \cite{JarrettSMIR},  the TDW-SMIR generator does not require calculating the spherical Bessel functions and the spherical Hankel functions at multiple wavenumbers (frequencies), as well as the inverse Fourier transform.

This section also compares the results of the SMIR generator and the TDW-SMIR generator for a source whose frequency invariant directivity is a cardioid. Suppose the cardioid's on-axis direction coincides with the positive $z^{(s')}$-axis, the amplitude of the spherical wave front is
\begin{align}
\label{Eq:cardioid_pattern}
    d(t, \theta^{(s')}, \phi^{(s')}) = &\bigg[\sqrt{\pi} Y_{0}^{0} (\theta^{(s')}, \phi^{(s')}) + \nonumber\\& \sqrt{\pi/3} \,Y_{1}^{0} (\theta^{(s')}, \phi^{(s')})\bigg]\delta(t).
\end{align}
The simulation considers a room of dimension [4, 6, 3] meters. The wall reflection coefficients are $\boldsymbol{\tilde{\beta}}=$[0.45, 0.7, 0.8, 0.5, 0.6, 0.75]. In Cartesian form, the source is located at $\mathbf{r_{s}^{(\varepsilon')}}=$ [1, 3.5, 2.1] meters, and the open spherical microphone array of radius 0.042 meters is centered at $\mathbf{r^{(\varepsilon')}}=$ [2.5, 3.5, 2.1] meters. The spherical microphone array has 32 microphones whose locations follow those of the Eigenmike \cite{mhacoustics}. Note it is assumed the on-axis direction of the cardioid points to the center of the spherical microphone array. Therefore, additional rotation to \eqref{Eq:cardioid_pattern} is needed. In total, 24 image sources are considered. The sampling frequency is 44.1 kHz. The SH truncation order of the observed signal is $N = 5$.

Figure \ref{Fig:ISM_cardioid_results} illustrates the impulse responses at microphone No. 17, which is at angular direction $(\theta^{(\rho')}, \phi^{(\rho')}) = (1.2043, \pi)$ rad. Note that the impulse responses are normalized so that the highest amplitude is 1. The results of the TDW-SMIR generator are similar to those of the SMIR generator. The oscillations (ringing) at each impulse are due to the band limit. 

The sampling method can be further improved. Equation \eqref{Eq:SH_coeff_obv_seq_gen} has finite duration due to the rectangular window $\Xi(t_{0}, \tau)$, which suggests classical impulse train sampling leads to aliasing \cite{Hahn2021sampling}. In \cite{Wang2023}, it was proposed that aliasing due to classical impulse train sampling could be reduced by applying a low-pass filter. Adding a low-pass filter introduces additional ringing in the simulation results. Therefore, the low-pass filter needs to be carefully designed. It is also possible to replace the rectangular window $\Xi(t_{0}, \tau)$ by other window functions with smaller sidelobe magnitude. Other advanced sampling and bandlimitation methods are covered in \cite{HahnAntiDerivPW} and \cite{HahnAntiDerivPS}. The incorporation of a proper aliasing reduction method should be considered in future work. 
\begin{figure}[t]
  \centering
  \includegraphics[trim = 30mm 42mm 10mm 45mm, clip, width = 0.9\columnwidth]{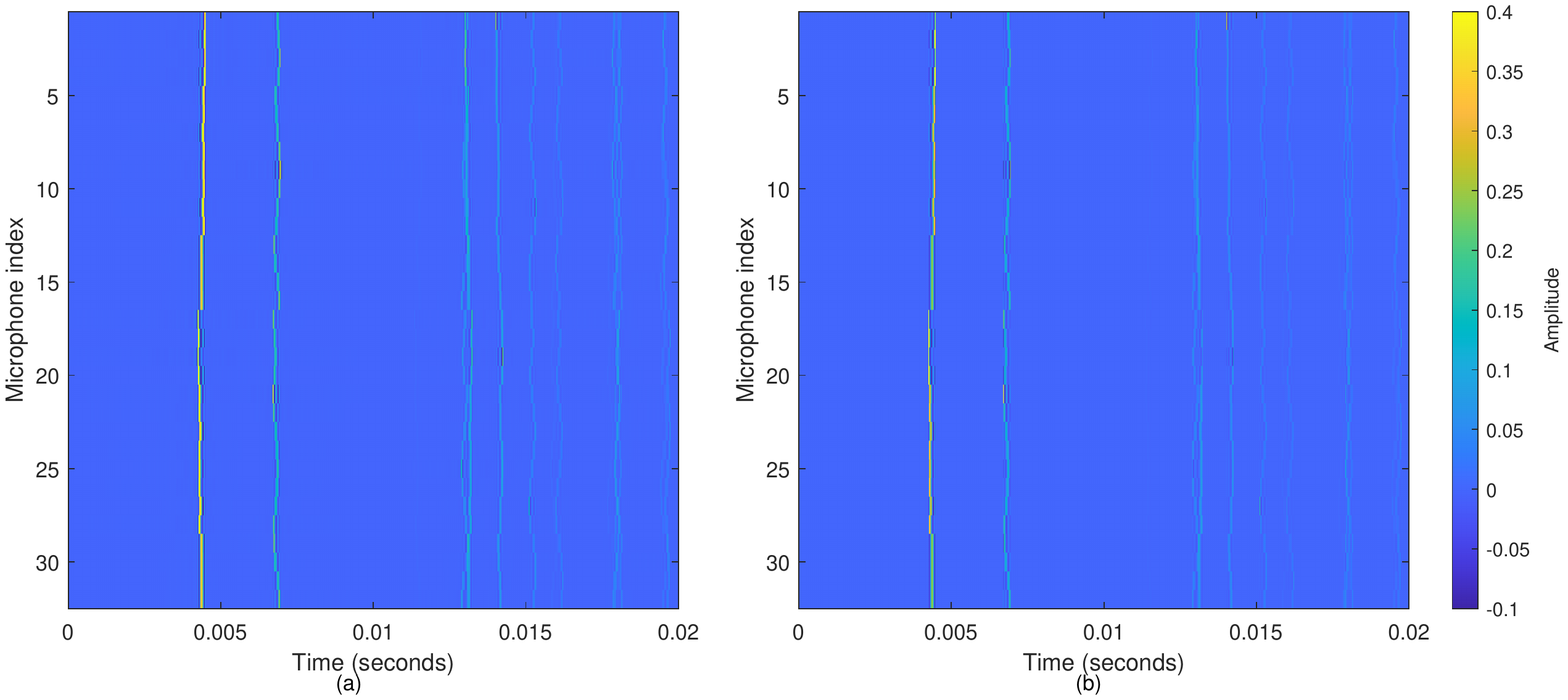}
  \vskip -3mm
  \caption{Comparison of the impulse responses at 32 microphones obtained using (a) the SMIR generator and (b) the TDW-SMIR generator. Only the first two 0.02 seconds are shown. }
  \vskip -3mm
  \label{Fig:ISM_cardioid_results_all_mics}
\end{figure}

Figure \ref{Fig:ISM_cardioid_results_all_mics} compares the impulse responses at 32 microphones. Only the direct path and first few dominant reflections are shown. Overall, the results of the the TDW-SMIR generator are similar to those of the SMIR generator. 

%%%%%%%%%%%%%%% Comparison with region-to-region RTF %%%%%%%%%%
\section{The TDW-SMIR generator with generalized source directivity}
\label{Sec:R2R_RTF_compare}
%%%%%%%%%%%%%%%%%%%%%%%%%%%%%%%%%%%%%%%%%%%%%%%%%%%%%%%%%%%%%%
This section presents the TDW-SMIR generator for a source with frequency dependent and arbitrarily shaped directivity. The generalization is achieved by using the loudspeaker directional impulse responses. When measuring the directional impulse responses of a loudspeaker, microphones are usually placed on a spherical surface surrounding the loudspeaker. In this paper, it is assumed the spherical surface is sufficiently large so that the measurements are in the far-field. Let $h(t, r^{(s)}, \theta^{(s)}, \phi^{(s)})$ denote the directional impulse responses of a loudspeaker measured on a spherical surface of radius $r^{(s)}$, 
\begin{equation}
\label{Eq:source_imp}
    h(t, r^{(s)}, \theta^{(s)}, \phi^{(s)}) = \int_{\tau} h(\tau, r^{(s)}, \theta^{(s)}, \phi^{(s)}) \delta(t - \tau) d\tau.
\end{equation}
The integral is the sequence of spherical wave fronts captured at successive time instances. At time instance $\tau$, the captured spherical wave front has directional dependent amplitude $h(\tau, r^{(s)}, \theta^{(s)}, \phi^{(s)})$. 

In the TDW-SMIR generator, it is assumed the spherical wave fronts only experience direction independent attenuation due to traveled distance. Although the assumption is only suitable for far-field, it allows the incorporation of loudspeaker directional impulse responses into the time-domain wideband ISM. Consider the source in \eqref{Eq:source_seq}, suppose 
\begin{equation}
\label{Eq:source_dir_match_imp}
    \widehat{d}(\tau, \theta^{(s)}, \phi^{(s)}) = 4\pi r^{(s)} h(\tau +r^{(s)}/c, r^{(s)}, \theta^{(s)}, \phi^{(s)}).
\end{equation}
Apply the assumption, the directional impulse responses of this source measured on the spherical surface of radius $r^{(s)}$ should follow exactly \eqref{Eq:source_imp}. Loudspeaker impulse responses at other points in the far-field can be extrapolated by applying time delay and distance-related attenuation to the source signal $d(t,  \theta^{(s)}, \phi^{(s)})$.

In the TDW-SMIR generator, it is assumed that the source emits a sequence of spherical wave fronts
\begin{align}
    d(t, \theta^{(s)}, \phi^{(s)}) &= \int_{\tau} \widehat{d}(\tau, \theta^{(s)}, \phi^{(s)}) \delta(t-\tau) d\tau \nonumber\\ &= \int_{\tau} h(\tau, r^{(s)}, \theta^{(s)}, \phi^{(s)}) \delta(t-\tau) d\tau.
\end{align}
Note that the time-aligning factor $r^{(s)}/c$ and the attenuation $4\pi r^{(s)}$ are ignored. They can be compensated by shifting the simulated room impulse responses to the left and applying additional amplitude scaling factor.

In this section, comparison is made with the region-to-region ISM generator (RISM generator) in \cite{PrasangaISM}. The RISM generator can be adapted for the simulation of room impulse responses captured by a spherical microphone array. Note that the RISM generator is a frequency-domain single band method. Here a simulation example is provided. The room is of dimension [4, 6, 3] meters and the wall reflection coefficients are $\boldsymbol{\tilde{\beta}}=$[0.45, 0.7, 0.8, 0.5, 0.6, 0.75]. In Cartesian form, the source is located at $\mathbf{r_{s}^{(\varepsilon')}}=$ [1, 3.5, 2.1] meters, and the open spherical microphone array of radius 0.042 meters is centered at $\mathbf{r^{(\varepsilon')}}=$ [2.5, 3.5, 2.1] meters. The microphone locations follow those of the Eigenmike\cite{mhacoustics}. The number of image sources is 24 and the sampling frequency is 44.1 kHz. The loudspeaker directional impulse responses of Genelec-8020A in \cite{KUG} are used. Details of the measurement setup are in \cite{Brandner2018}. Like \eqref{Eq:source_seq}, the loudspeaker directional impulse responses are truncated to $V = 5$. The signals observed by the spherical microphone array are truncated to $N = 5$.

\begin{figure}[t]
  \centering
  \includegraphics[trim = 40mm 80mm 40mm 90mm, clip, width = 0.9\columnwidth]{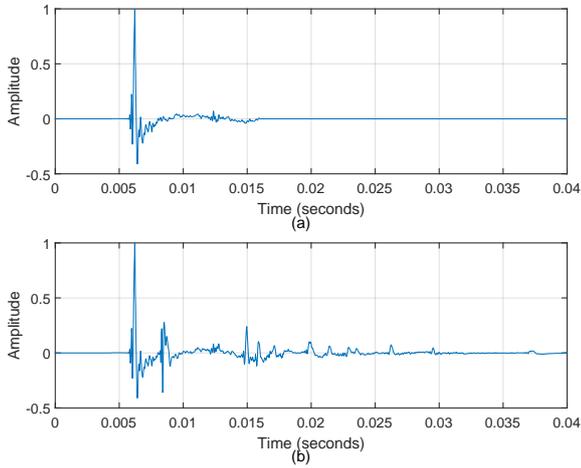}
  \vskip -3mm
  \caption{Room impulse responses at microphone No. 14 produced by the TDW-SMIR generator. (a) Anechoic condition. (b) Reverberant condition. }
  \vskip -3mm
  \label{Fig:ISM_DIR_mic_14}
\end{figure}

Figure \ref{Fig:ISM_DIR_mic_14} shows the room impulse responses obtained by the TDW-SMIR generator at microphone No. 14, which is at angular direction $(\theta^{(\rho')}, \phi^{(\rho')}) = (1.0123, \pi/2)$ rad. Figure 5(a) shows the anechoic condition (i.e., without the image sources) while Figure 5(b) shows the reverberant condition (i.e., with the image sources). Note that the impulse responses are normalized so that the highest amplitude is 1. It can be seen that the direct path in Figure 5(b) agrees with the signal in Figure 5(a). Figure 5(b) is essentially Figure 5(a) plus the signals generated by the image sources. 

\begin{figure}[t]
  \centering
  \includegraphics[trim = 40mm 80mm 40mm 90mm, clip, width = 0.9\columnwidth]{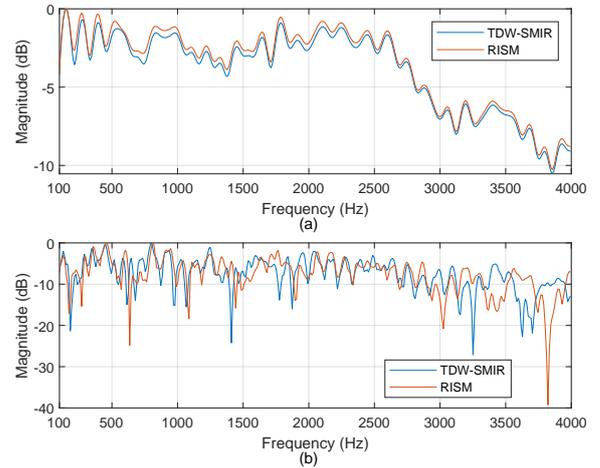}
  \vskip -3mm
  \caption{Room transfer function at microphone No. 14 produced by the TDW-SMIR generator and the RISM generator in (a) anechoic condition and (b) reverberant condition. }
  \vskip -3mm
  \label{Fig:ISM_DIR_freq_mic_14}
\end{figure}

Figure \ref{Fig:ISM_DIR_freq_mic_14} illustrates the room transfer functions captured at microphone No. 14. Note the room transfer functions are normalized so that the highest value is 0 dB. Figure \ref{Fig:ISM_DIR_freq_mic_14}(a) is the anechoic case and Figure \ref{Fig:ISM_DIR_freq_mic_14}(b) is the reverberant case. For the anechoic case, the room transfer functions obtained by the TDW-SMIR generator and the RISM generator are very similar. When there are room reflections, the room transfer functions produced by the two generators still agree in the low frequency, but start to deviate above 3 kHz. The reason for deviation could be insufficient number of SH coefficients.  

\begin{figure}[t]
  \centering
  \includegraphics[trim = 40mm 80mm 30mm 90mm, clip, width = 0.9\columnwidth]{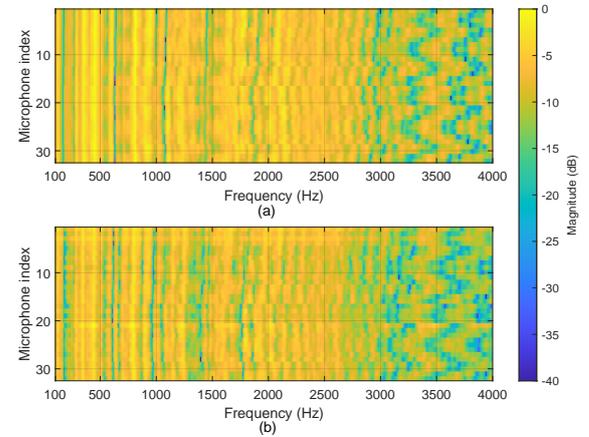}
  \vskip -3mm
  \caption{Room transfer functions at all 32 microphones produced by (a) the RISM generator and (b) the TDW-SMIR generator. }
  \vskip -3mm
  \label{Fig:ISM_DIR_freq_all_mics}
\end{figure}
Figure \ref{Fig:ISM_DIR_freq_all_mics} compares the room transfer functions captured at all 32 microphones in the reverberant condition. It can be seen that the TDW-SMIR generator and the RISM generator produce similar results below 3 kHz. 

\section{Conclusion}
This paper proposed the TDW-SMIR generator, which is a time-domain wideband ISM for simulating the room impulse response measured by an open spherical microphone array. To incorporate source directivities, the TDW-SMIR generator utilized measured loudspeaker directional impulse responses. The TDW-SMIR generator simulates the room impulse responses directly in the time-domain, which avoids the calculation of the inverse Fourier transforms, the spherical Bessel function and the spherical Hankel function often associated with frequency-domain single band method. Simulations showed the TDW-SMIR generator produced results that are similar to those of single band ISMs. Future work will investigate the perceptual effects of the room impulse responses obtained from the TDW-SMIR generator. The investigation will cover finding the optimal number of image sources to achieve adequate spatial impression, as well as incorporating a perceptually inspired method to generate late reverberation. The existence of sweeping echos due to rooms with perfect cuboid geometry will also be investigated.

\bibliographystyle{IEEEbib}
\bibliography{strings}
\vspace{12pt}

\end{document}